\documentclass[sn-mathphys,Numbered]{sn-jnl}

\usepackage{graphicx}%
\usepackage{multirow}%
\usepackage{amsmath,amssymb,amsfonts}%
\usepackage{amsthm}%
\usepackage{mathrsfs}%
\usepackage[title]{appendix}%
\usepackage{xcolor}%
\usepackage{textcomp}%
\usepackage{manyfoot}%
\usepackage{booktabs}%
\usepackage{algorithm}%
\usepackage{algorithmicx}%
\usepackage{algpseudocode}%
\usepackage{listings}%
\usepackage[normalem]{ulem}

\theoremstyle{thmstyleone}%
\theoremstyle{thmstyletwo}%

\theoremstyle{thmstylethree}%
\begin{document}

\title[RCAP of $^{11}$Be]{Investigating the rate of $^{10}$Be(n,$\gamma$)$^{11}$Be radiative capture reaction within the FRDWBA framework}


\author*[1]{\fnm{M.} \sur{Dan}}\email{mdan@rpr.amity.edu}

\author[2]{\sur{Shubhchintak}}\email{shubhchintak$\_$phy@pbi.ac.in}

\author[3]{\fnm{G.} \sur{Singh}}\email{gsingh@uni-mainz.de}

\author[4]{\fnm{V.} \sur{Choudhary}}\email{vchoudhary@ph.iitr.ac.in}

\author[5,6,7]{\fnm{Jagjit} \sur{Singh}}\email{jagjit.singh@manchester.ac.uk}

\affil[1]{\orgdiv{Amity School of Engineering and Technology}, \orgname{Amity University}, \orgaddress{\city{Raipur}, \postcode{493225}, \state{Chhattisgarh}, \country{India}}}

\affil[2]{\orgdiv{Department of Physics}, \orgname{Punjabi University Patiala}, \orgaddress{\city{Patiala}, \postcode{147002}, \state{Punjab}, \country{India}}}

\affil[3]{\orgdiv{Institut f{\"{u}}r Kernphysik}, \orgname{Johannes Gutenberg-Universit\"{a}t Mainz}, \orgaddress{\postcode{55099}, \state{Mainz}, \country{Germany}}}

\affil[4]{\orgdiv{Department of Physics}, \orgname{Indian Institute of Technology Roorkee}, \orgaddress{\city{Roorkee}, \postcode{247667}, \state{Uttarakhand}, \country{India}}}

\affil[5]{\orgdiv{Department of Physics and Astronomy}, \orgname{The University of Manchester}, \orgaddress{\city{Manchester},\postcode{M13 9PL}, \country{UK}}}
\affil[6]{\orgdiv{Department of Physics}, \orgname{Akal University}, \orgaddress{\city{Talwandi Sabo, Bathinda},\postcode{151302}, \state{Punjab}, \country{India}}}
\affil[7]{\orgdiv{Research Centre for Nuclear Physics (RCNP)}, \orgname{Osaka University}, \orgaddress{\city{Ibaraki},\postcode{567-0047}, \country{Japan}}}


\abstract{}

\abstract {This study examines the radiative capture of a neutron by $^{10}$Be using the Coulomb dissociation approach within the FRDWBA theory. We analyze the elastic Coulomb breakup of $^{11}$Be on a $^{208}$Pb target at 72\,MeV/A to determine the photodisintegration cross-section and radiative capture cross-section. Utilizing the Maxwell-averaged velocity distribution, we calculate the resulting radiative neutron capture reaction rate for the $^{10}$Be(n,$\gamma$)$^{11}$Be reaction. Comparative analyses are conducted with experimental data, theoretical results from direct radiative capture methods, and transfer reaction calculations. Additionally, we contrast our findings with the existing $^{10}$Be($\alpha$,$\gamma$)$^{14}$C reaction rate and conclude the dominance of neutron capture over $\alpha$ capture by $^{10}$Be.}




\maketitle

\section{Introduction}\label{sec1}


In the mid-eighties, Baur \textit{et al.} \cite{baur1986coulomb}, proposed the idea of inferring low energy radiative capture cross-sections from Coulomb breakup measurements, offering a successful alternative to direct approaches. In a typical Coulomb breakup or dissociation, virtual photons are exchanged between a weakly bound projectile and a heavy (high $Z$) target. This interaction due to the evolving Coulomb field of the target leads the projectile above its particle emission threshold, thereby initiating a breakup channel that is akin to the time-reversed reaction of the radiative capture \cite{baur1986coulomb,Baur1996,Baur2003,Bert88PR}. Further, their cross-sections are linked via the principle of detailed balance in the framework of perturbation theory \cite{baur1986coulomb,Baur2003}.
{The Coulomb dissociation (CD) method, now four decades old, stands as one of the most successful tools for probing novel exotic features such as the halo structure \cite{Tanihata1996,Riisager2013,Aumann2013} of drip line nuclei and their associated capture reactions of astrophysical significance. This success owes much to the progress made at radioactive ion beam (RIB) facilities, enabling the observation of halos\footnote{A halo is a composite system where one or two nucleons decouple from the core mean field of a nucleus, enabling their wave functions to penetrate the classically forbidden regions.} across a range of lighter to medium-mass nuclei.
Till date, the halo phenomenon has been observed in various nuclei via CD and interaction cross-section measurements, including one-neutron halos such as $^{11}$Be \cite{NakamuraPLB, Palit2003,fukuda2004coulomb}, $^{15,19}$C \cite{Nakamura1999,Nakamura2009a,DATTAPRAMANIK200363}, $^{31}$Ne \cite{Nakamura2009}, and $^{37}$Mg \cite{Kobayashi2014}, as well as two-neutron halos like $^{6}$He \cite{Aumann1999,Wang2002,SUN2021136072,FOrt2014,SFV22PLB,SFV16EPJ}, $^{11}$Li \cite{Tanihata1985,Nakamura11Li2006}, $^{14}$Be \cite{TANIHATA1988592,Labiche2001}, $^{17,19}$B \cite{Suzuki2002,Cook2020,Casal2020,Yama22PRC}, $^{22}$C \cite{Horiuchi2006,Tanaka2010,TOGANO2016412,JS19} and $^{29}$F \cite{SCH20PRC,BKT20PRL,FCH20CP,CSF20PRC}. Putative cases of neutron halos include $^{29}$Ne \cite{MDS21NPA}, $^{31}$F \cite{SSC22PRC}, $^{34}$Na \cite{SSC16PRC}, $^{39}$Na \cite{SINGH2024,Zhang2023}, $^{40}$Mg \cite{SINGH2024}, $^{42}$Al \cite{Zhang2023b} and, $^{62,72}$Ca \cite{Horiuchi2022}.}


The focus of the present study is on the CD of $^{11}$Be, which is the quintessential one-neutron halo and garners tremendous attention in the community due to its relatively simple two-body (core$+n$) structure. To date, the CD of $^{11}$Be has been studied experimentally at different energies: $520$\,MeV/A at GSI \cite{Palit2003}, $72$\,MeV/A \cite{NakamuraPLB}, and $69$\,MeV/A \cite{fukuda2004coulomb} at RIKEN. On the theoretical side, many efforts have been devoted to interpreting these measured data within different reaction frameworks \cite{CHATTERJEE2000,Chatterjee2002,Chatterjee2003,Chatterjee2007,RCPPNP,MOSCHINI2019,Singh2021,Moro12PRL,Calci16PRL,YC18PRC,Moro20PLB,PLM23PRC}. Despite the differences in these frameworks, almost all intricate calculations explain the measured relative energy spectra and deduced low-lying dipole strength distributions fairly well.

However, it is not only from the structure point of view that $^{11}$Be gains significance. The $^{10}$Be(n,$\gamma$)$^{11}$Be reaction is of paramount importance in understanding the astrophysical pathways and the formation of seed nuclei in reaction networks involving or leading up to the $r$-proccess \cite{Terasawa2001,Mengoni}. This reaction is also crucial for $^{12}$B formation via the ...$^{8}$Li(n,$\gamma$)$^{9}$Li($\beta^-$)$^{9}$Be(n,$\gamma$)$^{10}$Be(n,$\gamma$)$^{11}$Be($\beta^-$)$^{11}$B(n,$\gamma$)$^{12}$B... network, which is essential in the inhomogeneous Big Bang nucleosynthesis \cite{Kajino95NPA}. In fact, in the absence of one of the pathways for Carbon seed nuclei production that are supposedly relevant for the $r$-process, i.e., when the $^{10}$Be($\alpha$,$\gamma$)$^{14}$C rate is very low or negligible under given physical conditions, the $^{9}$Be(n,$\gamma$)$^{10}$Be(n,$\gamma$)$^{11}$Be(n,$\gamma$)$^{12}$Be($\beta^-$)$^{12}$B network takes over, producing $^{12}$B. The inclusion of these light neutron excess nuclei in reaction networks can allegedly affect the heavy element abundance patterns \cite{Terasawa2001}, and it is thus, absolutely imperative that constraints are put on each of the component reactions of these networks while respecting the microscopic structures of the nuclei involved \cite{GORIELY97AA}. We, therefore, analyze the rate of this $^{10}$Be(n,$\gamma$)$^{11}$Be radiative capture reaction through the lens of the CD of $^{11}$Be taking into account its halo and weakly bound nature.

We employ CD within the realm of the finite-range distorted-wave Born approximation (FRDWBA) theory, utilizing the virtual photon number method to indirectly calculate the photodisintegration cross-section. The FRDWBA theory, a fully quantum mechanical framework, has been extensively utilised in nuclear physics for various reactions, proving particularly elegant for identifying the isotope production flow towards the neutron drip line \cite{SSC17PRC,DSC19PRC,SCS17PRC}. 
In fact, past implementations have successfully demonstrated its efficacy in reaction rate calculations, notably in studies of neutron-rich medium mass nuclei \cite{DSC19PRC,SSC17PRC,SCS17PRC,SSC24FBS,Singh20JPCSr,CDC23arxiv,DSC22DAE}.
Using this method, we compute the relative energy spectrum for the elastic Coulomb breakup of $^{11}$Be on a $^{208}$Pb target at a beam energy of $72$\,MeV/A, subsequently deriving the photodisintegration cross-section from its dipole strength distribution and utilizing the detailed balance principle to determine the radiative capture cross-section. Then the radiative neutron capture reaction rate is easily obtained by incorporating the capture cross-section into the Maxwell-averaged velocity distribution. We then compare this (n,$\gamma$) reaction rate with experimental findings, as well as the rates obtained through direct radiative capture methods and transfer reaction calculations. Additionally, we collate it with the $^{10}$Be($\alpha$,$\gamma$)$^{14}$C reaction rate currently available from the literature. 

The next section of this paper explains the FRDWBA theory. Section 3 presents the results of our calculations, which include relative energy spectra, capture cross-section, and finally, the reaction rate. Section 4 summarises the findings of our research.

\section{Formalism}
\label{sec2}

Consider the reaction \textit{a} + \textit{t} $\rightarrow $ \textit{b} + \textit{c} + \textit{t}, where the dynamic Coulombic hostility of a heavy target nucleus `\textit{t}' forces a feebly bound projectile `\textit{a}' to break into substructures `$b$' and `$c$'. Here, we associate the projectile \textit{`a'} with $^{11}$Be, while $b$, $c$ and the target \textit{`t'} can then be thought of as $^{10}$Be, a neutron and $^{208}$Pb, respectively. 
Under such considerations, the triple differential cross-section for the reaction can be expressed as, 
\begin{eqnarray}
\dfrac{d^3\sigma}{dE_{b}d\Omega_{b}d\Omega_{c}} = \dfrac{2\pi}{\hbar v_{at}}\rho{(E_{b},\Omega_{b},\Omega_{c})}\sum_{\ell,m}|\beta_{\ell m}|^{2}.
\label{a2.1}
\end{eqnarray}

Here, the relative velocity between the \textit{a}-\textit{t} system in the entrance channel is denoted by $v_{at}$, while the three-body final state phase space factor is represented by $\rho{(E_{b},\Omega_{b},\Omega_{c})}$ \cite{Fuchs}. The relative orbital angular momenta between the projectile fragments are denoted by $\ell$, with $m$ being their projections. For a breakup process, the reduced transition amplitude $\beta_{\ell m}$ from the initial state to the final state in the post-form FRDWBA is given by \cite{RCPPNP,SC14NPA},

\begin{figure}[h]
\centering
\includegraphics[scale=0.35]{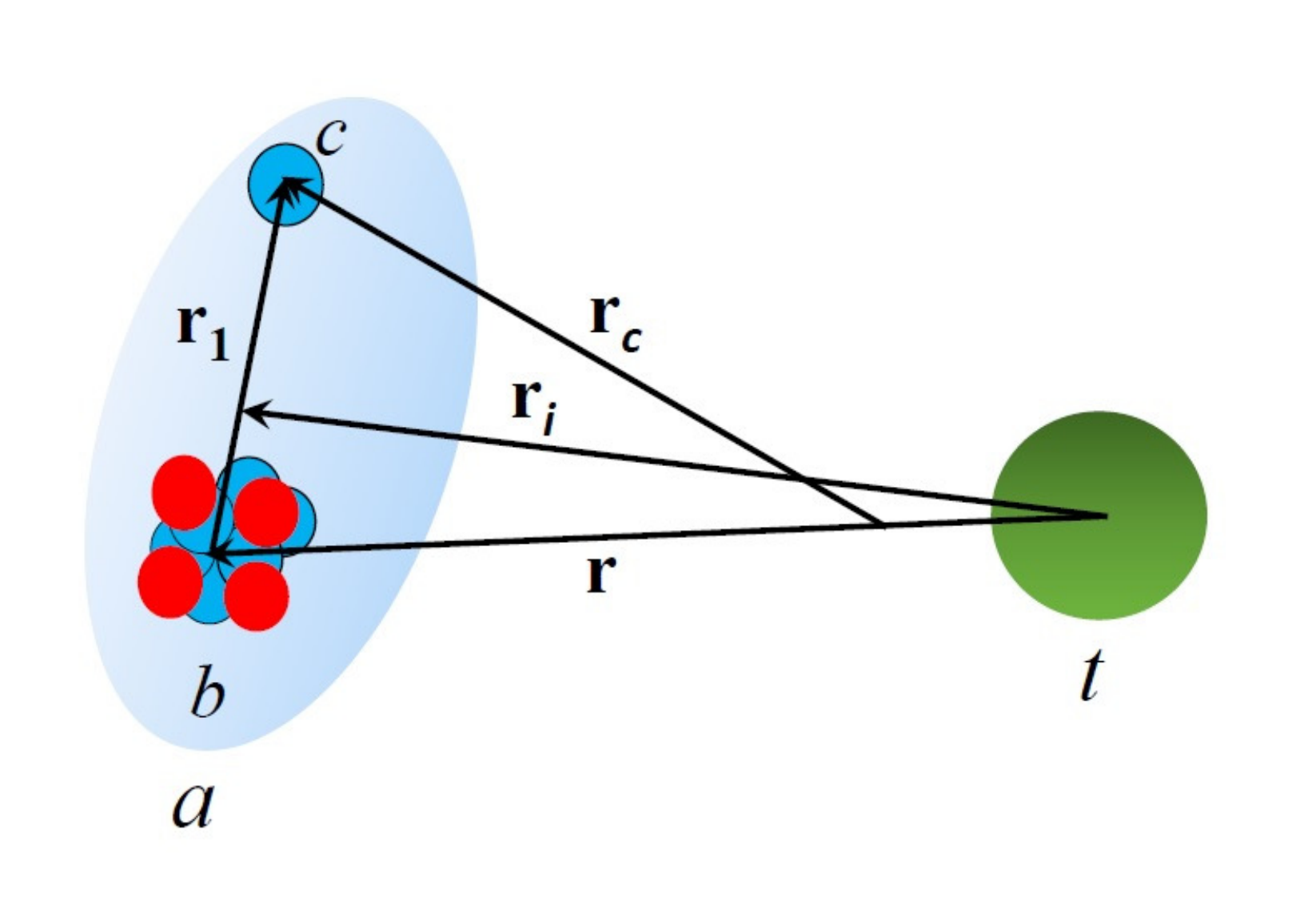}
\caption{The three-body Jacobi coordinate system. The corresponding position vectors are denoted by \textbf{r}'s. }
\label{structure}
\end{figure}

\begin{eqnarray}
\begin{aligned}
\beta_{\ell m} = \langle \chi_{b}^{(-)}(\textbf{r})\chi_{c}^{(-)}(\textbf{r}_{c})|V_{bc}(\textbf{r}_{1})| \phi_{a}^{lm}(\textbf{r}_{1})\chi_{a}^{(+)}(\textbf{r}_{i})\rangle.
\end{aligned}
\label{RedT}
\end{eqnarray}

With respect to the target, $\chi_{i}$'s (\textit{i = a, b, c}) in Eq. (\ref{RedT}) indicate the pure Coulomb distorted waves of the appropriate particles, which depend on Jacobi position vectors as well as wave vectors (the latter are not explicitly written for the purpose of brevity). The former can be visualized easily via Fig.~\ref{structure} to be related as \textbf{r} = \textbf{r}$_i$ - $\alpha$\textbf{r}$_1$ and \textbf{r}$_c$ = $\gamma$\textbf{r}$_1$ + $\delta$\textbf{r}$_i$, where the mass factors are $\alpha$ = m$_c$/(m$_c$ + m$_b$); $\delta$ = m$_t$/(m$_b$ + m$_t$); $\gamma$ = (1 - $\alpha\delta$), with m$_{i}$'s being the masses of the associated particles. The bound state wave function of the projectile is $\phi_{a}^{lm}$({\textbf{r}$_1$}), which is the only input to the FRDWBA in its post form. Meanwhile, $V_{bc}$(\textbf{r}$_1$) is the two-body bound state potential, which incorporates the finite range effects \cite{RCPPNP}. We take it here to be of the Woods-Saxon form whose depth was adjusted to reproduce the one neutron separation energy of the $^{11}$Be projectile. This is also where quadrupole deformation, if desired, can be inserted in the system \cite{SC14NPA,SSC16PRC,RCPPNP}. 

Once the triple differential cross-section is obtained, various other reaction observables, like the relative energy spectrum, parallel momentum distribution, and angular distribution can be evaluated by the appropriate integration of Eq.~\ref{a2.1}. Using the so obtained relative energy spectrum $d\sigma/dE_{rel}$ ($E_{rel}$ is the relative energy of $b$ and $c$ in the final channel), we can get the dipole strength distribution ${dB(E1)}/{dE_{rel}}$ as \cite{Manju19EPJ},

\begin{eqnarray}
\dfrac{dB(E1)}{dE_{rel}}\ = \left[ \dfrac{16\pi^3}{9\hbar c}{n_{E1}}\right ]^{-1}\dfrac{d\sigma}{dE_{rel}}\,,
\label{dbde}
\end{eqnarray}

where $n_{E1}$ is the virtual photon number for the electric dipole transition depending on the projectile and target combination. 
It can be calculated as,
\begin{eqnarray}
{\rm n}_{E 1}=\int_0^{\theta_{gz}}\frac{d{\rm n}_{E 1}}{d\Omega_{at}}d\Omega_{at},
\label{vpn}
\end{eqnarray}
where $\theta_{gz}$ is the grazing angle for the scattering and $\dfrac{d{\rm n}_{E 1}}{d\Omega_{at}}$ is the equivalent photon number per unit solid angle for electric dipole transitions. For more details about the virtual photon number one is referred to Ref. \cite{baur1986coulomb}.

It is here that the importance of obtaining the right relative energy spectrum as an observable shines through. Once known, it can further, for a dipole dominated pure Coulomb breakup, be related to the photo-disintegration cross-section $\sigma_{(\gamma,n)}$, as illustrated in \cite{baur1986coulomb}, via

\begin{equation}
\begin{aligned}
\dfrac{d\sigma}{dE_{rel}} = \dfrac{1}{E_{\gamma}}{\sigma_{(\gamma,n)}}{n_{E1}}.
\label{rel_eng}
\end{aligned}
\end{equation}

In the above expression, ${E_{\gamma}}$ = $E_{rel}$ + $S_{n}$; $S_{n}$ being the energy needed to separate one neutron from the projectile nucleus. 
Alternatively, one can also find the photo-disintegration cross-section directly from the dipole strength distribution using the relation 
\begin{equation}
\begin{aligned}
\sigma_{(\gamma, n)} = \dfrac{16\pi^3}{9\hbar c}{E_{\gamma}}\dfrac{dB(E1)}{dE_{rel}}.
\label{dbde3}
\end{aligned}
\end{equation}

Interestingly, the radiative capture cross-section, $\sigma_{(n,\gamma)}$ is just the time-reversed counterpart of the photo-disintegration cross-section and thus, can be easily computed using the well recognized principle of detailed balance if the latter is known \cite{baur1986coulomb}: 

\begin{equation}
\begin{aligned}
\sigma_{(n,\gamma)} = \dfrac{2(2\textit{j}_a+1)}{(2\textit{j}_b+1)(2\textit{j}_c+1)}\dfrac{\textit{k}_\gamma^2}{\textit{k}^2}{\sigma_{(\gamma,n)}},
\label{rel_eng2}
\end{aligned}
\end{equation}

where, $\textit{k}_\gamma$ is the wave number of the photon, while $\textit{k}^2 = \dfrac{2\mu_{bc} E_{rel}}{\hbar^2}$, with $\mu_{bc}$ being the reduced mass of the \textit{b}-\textit{c} system. $j_{a}$, $j_{b}$, and $j_{c}$ denote the total angular momenta of particles $a$, $b$, and $c$ respectively.

Then the reaction rate per particle pair per mole, for the radiative capture reaction $\textit{b} + \textit{c}  \longrightarrow  \textit{a} + \gamma$ is,

\begin{equation}
\begin{aligned}
R=N_A\langle\sigma_{(n,\gamma)}\textit{v}\rangle, 
\label{rateastro}
\end{aligned}
\end{equation}

where $N_A$ is the Avogadro number and the cross-section for a single target nucleus at a relative velocity $v$ is $\sigma$. The reaction rate per particle pair is $\langle\sigma_{(n,\gamma)}\textit{v}\rangle$, which is averaged over the Maxwell-Boltzmann velocity distribution~\cite{rolfs1988cauldrons}. In units of $\textrm{cm}^3$ $\textrm{s}^{-1}$ it is defined as,

\begin{eqnarray}
\langle\sigma_{(n,\gamma)}\textit{v}\rangle = \sqrt\frac{8}{(\pi\mu_{bc}(k_BT_{s})^3)} \Bigg\{\int_{0}^{\infty}dE_{rel}\sigma_{(n,\gamma)}(E_{rel})  \times E_{rel}e^{(-E_{rel}/k_BT_{s})}\Bigg\},
\label{rateeqn}
\end{eqnarray}

where $T_{s}$ is the stellar temperature\footnote{Stellar temperatures are usually expressed in terms of $T_3$, $T_6$, $T_9$, etc., where the suffix refers to the power of $10$. Hence, these temperatures would correspond to $10^3$\,K, $10^6$\,K, $10^9$\,K and so on.}, and $k_B$ is the Boltzmann constant. More details about the formalism can be found in Refs. \cite{SSC17PRC,DSC19PRC,SCS17PRC,SSC24FBS,Singh20JPCSr}. 

A word here is in order about the necessity of such an indirect approach of computing the observables of the reverse reaction and using the principle of detailed balance to obtain the required cross-sections. This is because it has been shown \cite{SSC17PRC,SCS17PRC,DSC19PRC,SSC24FBS} that the integrand in Eq. (\ref{rateeqn}) above is substantial only at very low relative energies. On the other hand, even a stellar temperature of $T_9$ = 1 roughly equates to a centre of mass energy of 0.1\,MeV and although there has been a tremendous advancement in the experimental techniques over the last few decades throughout the world, the typically low values of cross-sections make direct measurements at such low relative energies extremely tedious, thereby necessitating indirect approaches. Several indirect techniques have been developed to tackle the problem \cite{BSM16JPCSr,CRA14PRC,TBC14RPP} and CD is just one of them. With this view of the formalism, we now turn to the results.

\section{Results and discussions}\label{sec2}

\begin{figure}[!h]
\centering\includegraphics[width=2.8in]{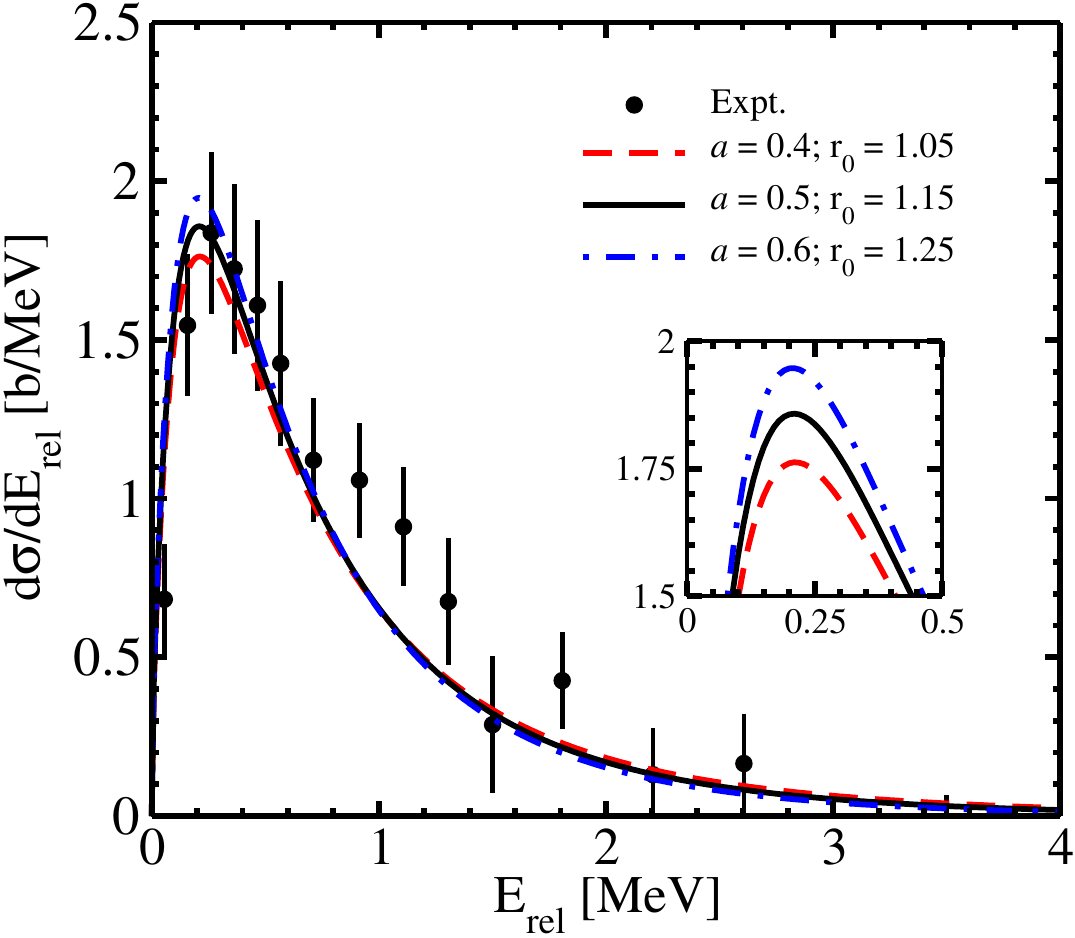}
	\caption{The relative energy spectra in the elastic Coulomb breakup of $^{11}$Be on $^{208}$Pb at $72$\,MeV/A calculated using the FRDWBA for various combinations of radius and diffuseness parameters. The solid circles show the experimental breakup data from Ref. \cite{NakamuraPLB}. The inset shows the peaks of the spectra up to a relative energy of $0.5$\,MeV.}
\label{rel}
\end{figure}

$^{11}$Be is considered to be the archetypical one-neutron halo and with a separation energy of $S_n$ = $0.504$\,MeV and a spin-parity of $1/2^+$ \cite{Wang21CPC}, it manifests beautifully the phenomenon of shell inversion. 
We take these properties into account and begin our quest for the rate of the $^{10}$Be(n,$\gamma$)$^{11}$Be reaction with the calculation of the elastic Coulomb breakup of $^{11}$Be on a Pb target at $72$\,MeV/A using the FRDWBA theory. We first calculate the relative energy spectrum, which is an important ingredient in the procedure of obtaining capture cross-section from the Coulomb breakup. The results are shown in Fig.~\ref{rel}. Keeping the one-neutron separation energy fixed at $0.504$\,MeV, we analyze the breakup using \textit{ground state} wave function of $^{11}$Be obtained with three different sets of WS parameters, shown by the red dashed ($r_0$ = $1.05$\,fm, a= $0.4$\,fm), black solid ($r_0$ = $1.05$\,fm, a = $0.5$\,fm) and the blue dot-dashed lines ($r_0$ = $1.25$\,fm, a= $0.6$\,fm), respectively. 

We observe that each of them can, within the error bars, generate curves that replicate the peak position of the experimental relative energy spectrum while producing almost indiscernible results in the asymptotic regions. For our purpose of subsequent calculations, however, out of the three combinations, we stick to a diffuseness parameter of $0.5$\,fm and a radius parameter of $1.15$\,fm as they are consistent with the literature \cite{CHATTERJEE2000}.
One may note that the spectroscopic factor (SF) used for this comparison is taken as 1 \cite{Aumann2013, NakamuraPLB, fukuda2004coulomb}, which reproduces the peak of the spectrum nicely. We observe that with $8-20\%$ change in potential parameters $r_0$ and $a_0$, the SF changes only by $5\%$, which still lies within the range available in the existing literature.

It is also evident that there is a slight deviation of the experimental data from the theoretical curves around $1$\,MeV. This is because the experimental data shown here is for the total breakup cross-section, having contributions from Coulomb, nuclear as well as from Coulomb nuclear interference (CNI) terms, while the theoretical curve describes only the pure Coulomb breakup. It must be noted that in Ref.~\cite{NakamuraPLB}, it is mentioned that this nuclear contribution, for $^{11}$Be breaking up on $^{208}$Pb, is estimated to be only about $10\%$ at the peak position, and was obtained by scaling the nuclear contribution on a C-target. This was not, however, the case in the theoretical study of Ref.~\cite{Chatterjee2003} (which carried out the analysis within the same model as in the present work), where the nuclear contribution at the peak was found to be negligible and only started to rise at higher relative energies. In fact, extracting the total one neutron removal cross-section ($\sigma_{-1n}$), which can be obtained by integrating the relative energy spectrum, we found $\sigma_{-1n}$ to be $1.76$\,b from the pure Coulomb breakup, while the corresponding experimental value including nuclear effects is $1.8\pm0.4$\,b~\cite{NakamuraPLB}.

We want to stress that we are primarily interested in the radiative neutron capture reaction rate of $^{10}$Be, for which at relatively low temperature ($T_9$), almost all the contribution arises from the low-energy domain~\cite{SCS17PRC}, and where nuclear contributions can be safely ignored as they contribute minimally in this energy region~\cite{Chatterjee2003}.
Nevertheless, to avoid any possible inconsistency in the results due to possible nuclear and CNI effects in the breakup of $^{11}$Be on $^{208}$Pb, we now use the pure Coulomb breakup data of the $dB(E1)/dE_{rel}$ spectrum (adopted from Fig. 1(b) of Ref.~\cite{NakamuraPLB}) for further comparisons and calculations of capture cross-section and consequently, the reaction rate.


\begin{figure}[!h]
\centering\includegraphics[width=2.8in]{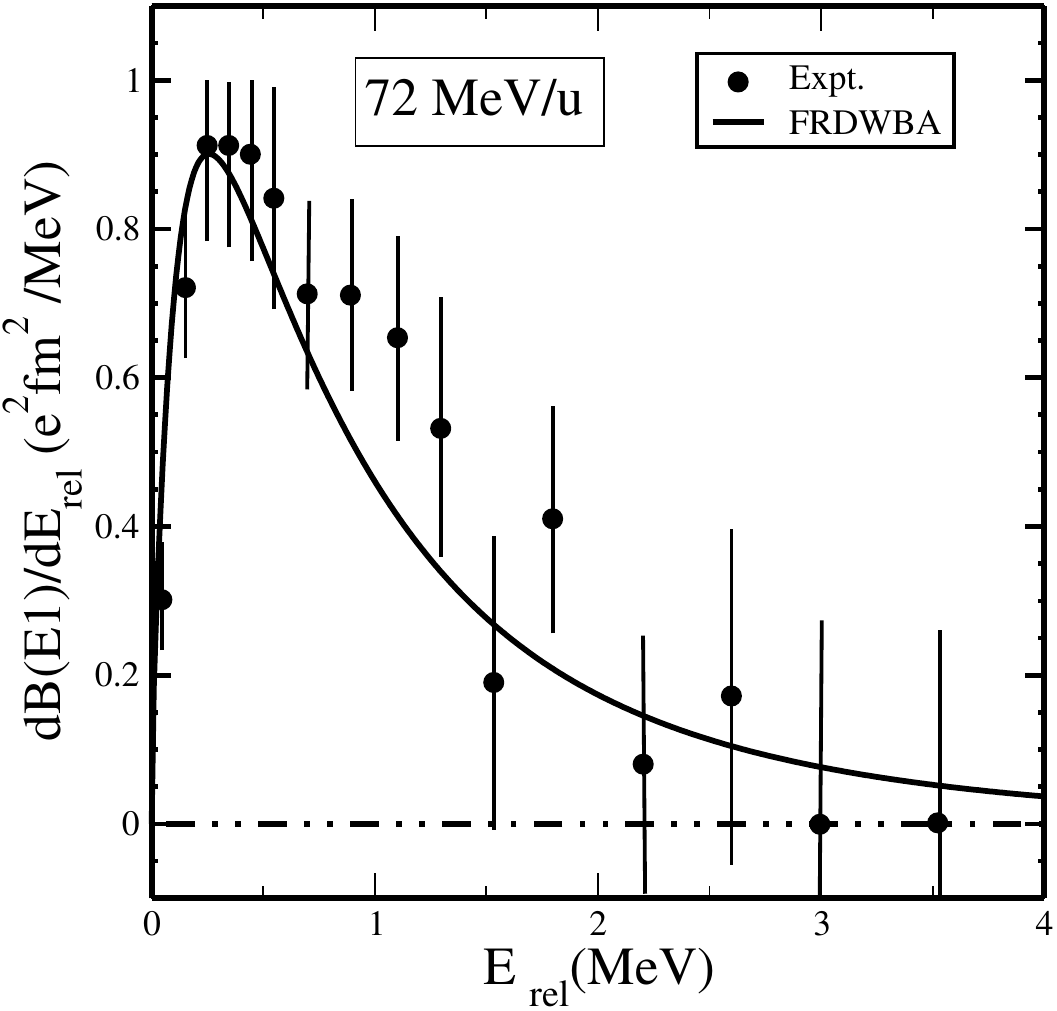}
	\caption{The dipole strength distribution in the elastic Coulomb breakup of $^{11}$Be on $^{208}$Pb at $72$\,MeV/A calculated using the FRDWBA. The solid circles show the experimental data from Ref. \cite{NakamuraPLB}.}
\label{dbdefig}
\end{figure} 

Hence, utilizing Eq.~(\ref{dbde}), we calculate the dipole strength distribution as shown in Fig.~\ref{dbdefig}, which also agrees very well with the pure Coulombic experimental data taken from Ref. \cite{NakamuraPLB}. The virtual photon number required for these calculations is computed by integrating Eq. (\ref{vpn}) up to an upper limit of the grazing angle that is taken at $2.3^{\circ}$ to match the experimental peak cross-sections \cite{NakamuraPLB}. Note that the experimentally reported total B(E1) value for $^{11}$Be is $1.3\pm0.3$\,$e^{2}fm^{2}$ \cite{NakamuraPLB}, while the theoretical value gained by integrating the curve in Fig.~\ref{dbdefig} is $1.17$\,$e^{2}fm^{2}$. Evidently, the two values match very well. These results suggest that, at low energies, the enhanced dipole strength is caused by the nucleus going above the breakup threshold and not by any soft dipole resonance. The authors of Ref.~\cite{fukuda2004coulomb} also came to similar conclusions.

\begin{figure}[!h]
 \centering\includegraphics[width=2.8in]{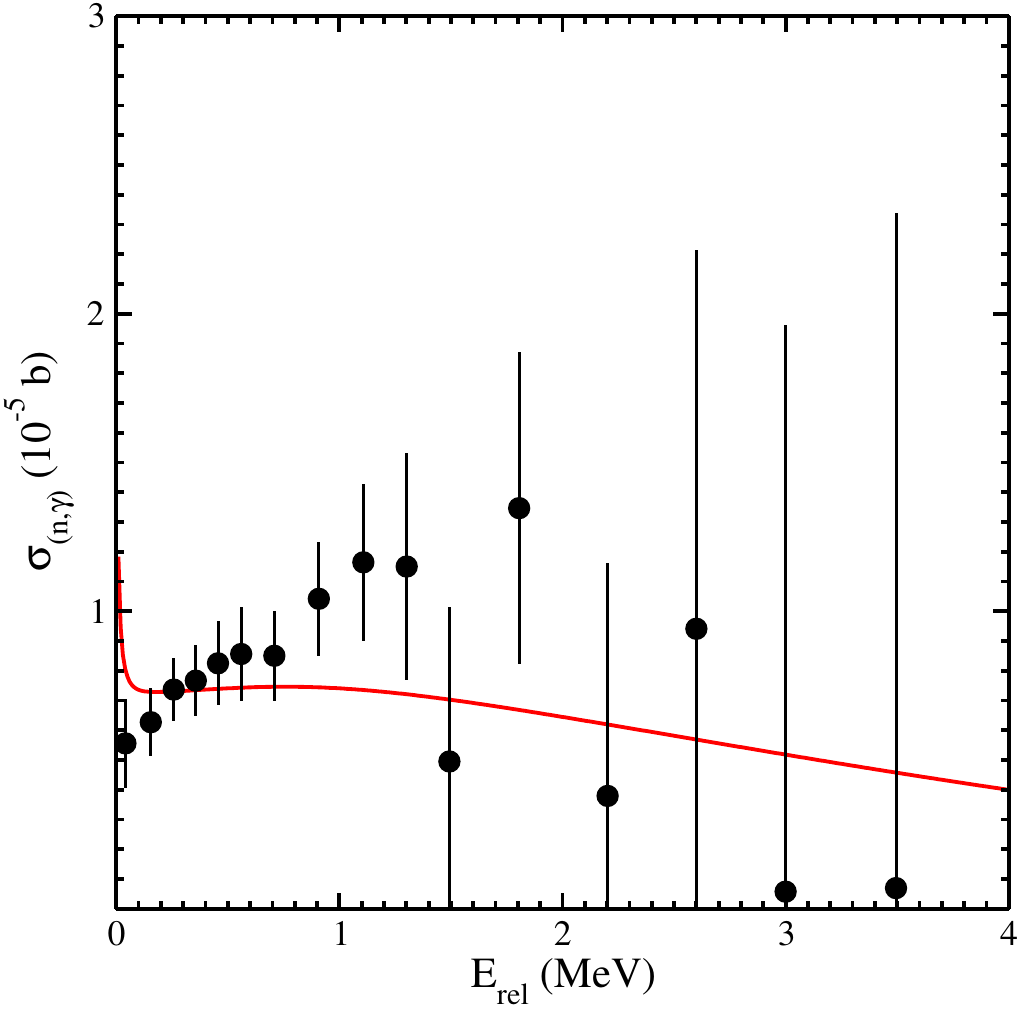}
	\caption{Radiative neutron capture cross-section of the $^{10}$Be($n,\gamma$)$^{11}$Be reaction. The experimental data are propagated from Ref.~\cite{NakamuraPLB}.}
\label{cap}
\end{figure}

Next, we calculate the photo-disintegration cross-section using Eq.~(\ref{dbde3}). To compare this FRDWBA result with the experiment, we also deduced the experimental photo-disintegration cross-sections, which are obtained from $dB(E1)/dE_{rel}$ data that is pure Coulomb in nature. From the photo-disintegration cross-section, we calculate the radiative neutron capture cross-section utilizing Eq.~(\ref{rel_eng2}).
In Fig. \ref{cap}, we plot our calculated radiative neutron capture cross-section and compare it with the experimental data propagated from the measured dipole strength distribution of Ref. \cite{NakamuraPLB} using the same procedure and conditions as adopted for the theoretical calculations. A similar procedure was also adopted in Ref. \cite{Mengoni} to deduce the capture cross-sections corresponding to the measured breakup data.

Our calculated cross-section shows a fair agreement with the deduced data. A small kink at very low energies ($\leq 100$ keV) in the theoretical capture cross-section is seen. This arises due to the factor $k^3_{\gamma}/k^2$ (from Eqs. (\ref{rel_eng}) and (\ref{rel_eng2})), where at low energies the denominator approaches zero while the numerator remains finite. This was also explored in detail in a study on radiative neutron capture by $^{33}$Na \cite{SSC17PRC}.

Using our calculated neutron capture cross-section we calculate the reaction rate of $^{10}$Be(n,$\gamma$)$^{11}$Be described by Eqs. (\ref{rateastro}-\ref{rateeqn}). This is the first attempt, within the `post form' reaction theory, to find this reaction rate from Coulomb dissociation. As averred in the introduction, $^{10}$Be(n,$\gamma$)$^{11}$Be is an important reaction in the network of inhomogeneous Big Bang nucleosynthesis. Moreover, $^{10}$Be(n,$\gamma$)$^{11}$Be reaction also has influence in determining the abundance of mass numbers A = 11-12 and higher. In Ref. \cite{Terasawa2001}, Terasawa \textit{et al.} suggest an $r$-process in the low mass region, where depending upon the temperature and pressure, they discuss the influence of neutron and $\alpha$ capture processes on the abundance of seed nuclei. Indeed, the radiative $\alpha$ capture reaction on $^{10}$Be could lead to the formation of $^{14}$C, whereas the dominance of successive neutron captures on $^{10}$Be would form $^{12}$Be which after $\beta$-decay forms the $^{12}$B \cite{Terasawa2001, SSC24FBS}. Therefore, it becomes interesting to compare the neutron and $\alpha$ capture rates on $^{10}$Be over a wide temperature range.



\begin{figure}[!h]
\centering\includegraphics[width=3.5in]{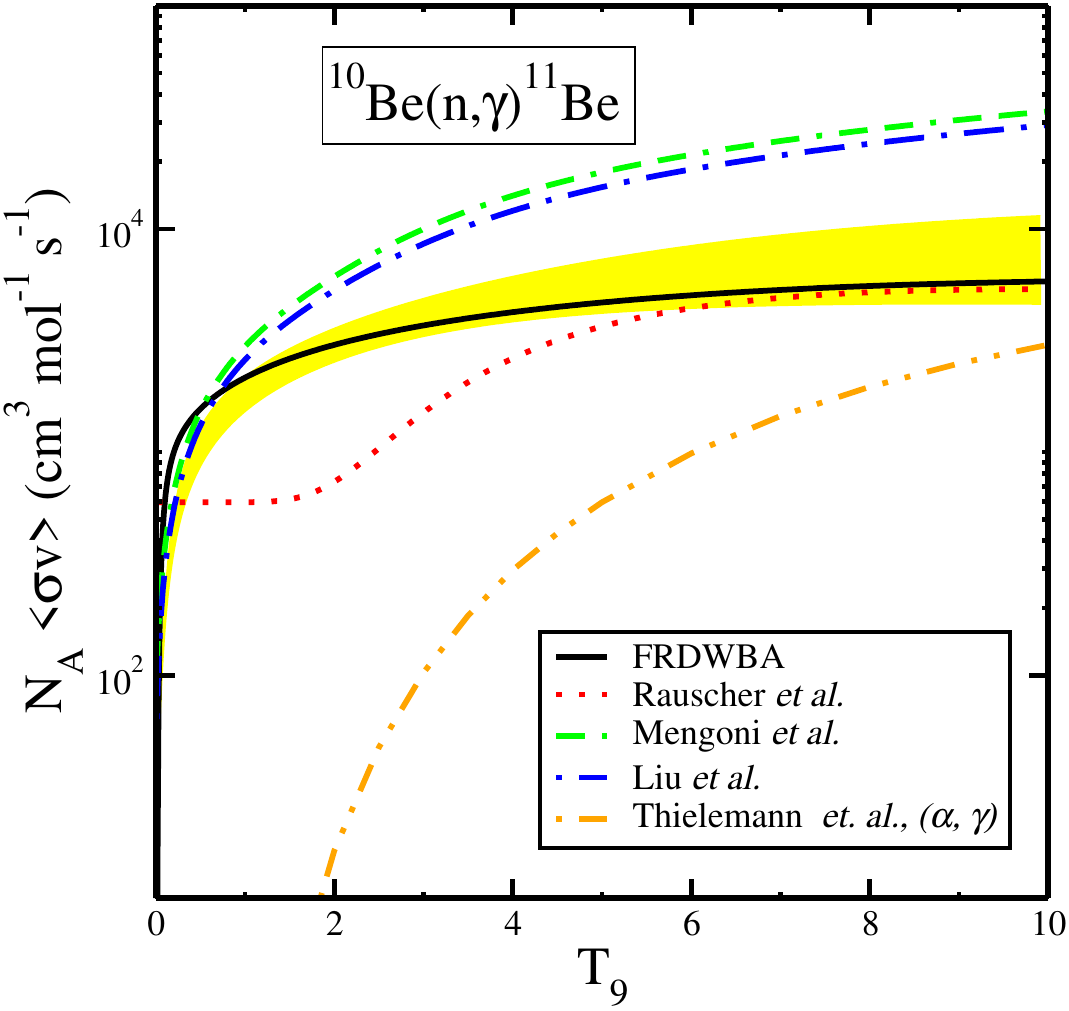}
	\caption{Reaction rate for the $^{10}$Be$(n,\gamma)^{11}$Be reaction plotted as a function of temperature. The solid line corresponds to FRDWBA calculation while the experimental data developed from Ref. \cite{NakamuraPLB} is shown by the shaded band. The dotted (Rauscher \textit{et al.} \cite{Rauscher}), dot-dashed (Liu \textit{et al.} \cite{liu}) and the double-dash-dotted (Mengoni \textit{et al.} \cite{Mengoni}) lines represent the reaction rates from some of the previous estimates. The double-dot-dashed curve shows the $\alpha$ capture rate extracted from Thielemann \textit{et al.} \cite{Thielemann}.}
\label{rate11be}
\end{figure}


In Fig. \ref{rate11be}, we plot our calculated rate for the $^{10}$Be$(n,\gamma)^{11}$Be radiative capture reaction (solid line) and compare it with those from some previous estimates.

The experimental band, shown in yellow, is generated by plugging in the experimentally deduced capture cross-section (shown in Fig. \ref{cap}), taking into account the error bars, into Eq. (\ref{rateastro}). A comparison of this experimental band with some of the other results can also be made. It is noteworthy that while the statistical estimates of the dotted curve (taken from Ref. \cite{Rauscher}) do not take into account the halo nature and parity inversion in $^{11}$Be, the rate from the double-dash-dotted curve (of Ref. \cite{Mengoni}) goes higher at higher temperatures because it also considers the transitions to its first excited state under a direct radiative capture model. Same is the case with the dot-dashed curve \cite{liu}. However, note that the experimental data was measured only for the ground state of $^{11}$Be \cite{NakamuraPLB}. 
Analogous rate calculations were also done in Ref. \cite{DUBOVICHENKO201991}, where the authors adopted a modified potential cluster model framework categorising states based on Young's tableaux. 

On the other hand, the FRDWBA results match fairly well with the experimental band throughout the temperature range shown here. This highlights the importance of proper considerations of structure inputs like spin and binding energies, especially for exotic nuclei. Finally, Fig. \ref{rate11be} also shows the $\alpha$ capture reaction rate of the $^{10}$Be($\alpha$,$\gamma$)$^{14}$C reaction (extracted from Ref. \cite{Thielemann}) via the double-dot-dashed curve. We observe that the (n,$\gamma$) reaction dominates over ($\alpha$,$\gamma$) reaction throughout the temperature range of $T_9 = 1.0 - 10$. This could potentially lead to the formation of $^{12}$B and thereby affect the abundance patterns of the Carbon isotopes. However, one must take this with a pinch of salt as there is a need to constrain the $^{10}$Be($\alpha$,$\gamma$)$^{14}$C reaction experimentally as well as theoretically for a definite conclusion about the temperature range where (n,$\gamma$) or ($\alpha$,$\gamma$) will dominate.

\section{Conclusions}\label{sec13}

To summarise, we have computed the radiative neutron capture rate for the $^{10}$Be(n,$\gamma$)$^{11}$Be reaction, for which we adopted the elastic Coulomb breakup of $^{11}$Be on $^{208}$Pb at 72\,MeV/A as an indirect approach. The $^{10}$Be(n,$\gamma$)$^{11}$Be reaction becomes crucial to the formation of $^{12}$B which is a vital step in the production of Carbon isotopes \cite{SSC24FBS}. The possible presence of nuclear effects in the relative energy spectrum of the breakup of $^{11}$Be on $^{208}$Pb prompted us to use the dipole distribution (for which the deduced data is pure Coulomb in nature) and subsequently compute the capture cross-section and reaction rate from our CD results. We applied CD via the FRDWBA theory, which is a fully quantum mechanical method that includes the entire non-resonant continuum of the projectile and fragment-target interactions up to all orders. 

Here, it is important to remember the one-step approximations made in the construction of the theory of FRDWBA, but at such a high beam energy it is reasonably justified. Also, the reaction under consideration must be dominated by a single multipole and free from any resonance. Furthermore, in the post-form FRDWBA, the initial state wave function is quite simplified which, indeed, is a plus point of the theory. 
No doubt, the occurrence of nuclear breakup and the interference effects between Coulomb and nuclear forces can always be witnessed in a reaction. Nevertheless, this issue can be significantly reduced by ensuring that the breakup occurs on a target with a high atomic mass and all measurements are conducted at forward angles for a Coulomb dominant breakup. While using CD as an indirect method in nuclear astrophysics, the nuclear breakup is not of interest, however, for proper structure studies one can use the extended version of the theory given in Ref. \cite{Chatterjee2003}, which treats Coulomb, nuclear and their interference on an equal footing.
In fact, one can also use other approaches integrated within the FRDWBA method to describe a halo nucleus like $^{11}$Be (cf. Ref. \cite{MDEPJA11BE}). 

We observe that our results obtained taking into account the proper structure of $^{11}$Be ground state match quite well with the experimentally derived results. Comparison with an $\alpha$-capture on $^{10}$Be showed that the neutron capture rate was much higher under the given physical conditions, which could push the network production of $^{12}$B and ultimately, Carbon isotopes. It is known, however, that even core excited states can contribute towards reaction rates at higher temperatures \cite{Singh20JPCSr,RAT16PRC,DSC22DAE} and $^{11}$Be is known to have significant core excitation effects \cite{Moro12PRL}. Therefore, it would be interesting to explore this domain further, but we leave it for the future.


\backmatter



\bmhead{Acknowledgments}

[S] acknowledges SERB, DST, India for a Ramanujan Fellowship (RJF/2021/000176). [GS] would like to acknowledge PRISMA+ (Precision Physics, Fundamental Interactions and Structure of Matter) Cluster of Excellence, Johannes Gutenberg University Mainz. This work was supported by the UK Science and Technology Funding Council [grant number
ST/V001116/1] (JS).

\bmhead{Conflict of Interest}
The authors declare that they have no conflict of interest.

\bmhead{Data Availability Statement}
This manuscript has no associated data or the data will not be deposited.



\bibliography{sn-bibliography}

\end{document}